\documentclass[aps,preprint]{revtex4}
\usepackage{amsfonts}
\usepackage{amsmath}
\usepackage{amssymb}
\usepackage{graphicx}
\usepackage{color}
\usepackage{hyperref}

\providecommand{\U}[1]{\protect\rule{.1in}{.1in}}

\begin{document}

\preprint{}
\title[ ]{Quantum renormalization group of XY model in two-dimensions}
\author{M. Usman$^{1}$}
\author{Asif Ilyas$^{2}$}
\author{Khalid Khan$^{1}$}
\email{kk@qau.edu.pk}
\affiliation{$^{1}$Department of Physics, Quaid-i-Azam University, Islamabad}
\affiliation{$^{2}$ Kohat University of Science and Technology, Kohat, KPK}
\date{May 12 2015}

\begin{abstract}
We investigate entanglement and quantum phase transition (QPT) in a
two-dimensional Heisenberg anisotropic spin-1/2 XY model, using quantum
renormalization group method (QRG) on a square lattice of $N\times N$ sites.
The entanglement through geometric average of concurrences is calculated
after each step of the QRG. We show that the concurrence achieves a non zero
value at the critical point more rapidly as compared to one-dimensional
case. The relationship between the entanglement and the quantum phase
transition is studied. The evolution of entanglement develops two saturated
values corresponding to two different phases. We compute the first
derivative of the concurrence, which is found to be discontinuous at the
critical point $\gamma=0$, and indicates a second-order phase transition in
the spin system. Further, the scaling behaviour of the system is
investigated by computing the first derivative of the concurrence in terms
of the system size.
\end{abstract}

\maketitle

\section{INTRODUCTION}

In quantum systems, entanglement is a resource that reveals the difference
between classical and quantum physics \cite{xydm1}. Its role has been
considered very vital to implement the quantum information tasks in
innovative ways like in quantum computations, quantum cryptography and
quantum teleportation etc. \cite{NC 2000}. In ecent years, the study of
entanglement in strongly correlated systems have attracted much more
attention \cite{vedral intro(1), vedral intro(3)} because it can describe
not only\ the information processing through correlation of spins \cite%
{vedral intro(7)} but also the critical phenomenon, quantum phase transition
(QPT) \cite{vedral intro(2)}. Therefore, the quantum entanglement is
considered as the common ground between the quantum information theory (QIT)
and the condensed matter physics \cite{xydm8, Langari xxz 1}. Recently, much
efforts have been devoted to the study of Heisenberg spin models,
especially, one dimensional spin models are the most explored area of
research, as these systems are exactly solvable and give quantitative
results \cite{xy3, xy4, xy5, ising, xxz, xxzdm, xy, xydm, vedral spin1/2(1),
vedral spin 1/2(2)}.

\textit{In higher dimensions, almost all the analysis of entanglement and
the QPT were made through numerical simulations} \cite{NS1, NS2}. \textit{%
Whereas the study of the phase diagram was also carried out} \cite{PD1, PD2,
PD3}. \textit{Using Monte Carlo simulations, concurrence was considered as
an entanglement measure in two-dimensional XY and XXZ models} \cite{NS1, NS2}%
. \textit{In the d-dimension pair wise entanglement was studied in XXZ model 
\cite{NS3, NS4} \ Concurrence was used to calculate the quantum entanglement
in the spin}$-1/2$\textit{\ ladder with four spins ring exchange by exact
diagonalization method} \cite{NS5}.

The density-matrix renormalization group method is a leading numerical
technique useful in exploring ground state properties for many body
interactions in lower dimensions \cite{xydm16, xydm17, xydm18, xydm19}.
Alongside, quantum renormalization group method (QRG) is another technique
which deals with large size systems analytically. At low temperatures
behavior of the spin systems effect their quantum nature due to quantum
fluctuations. At these temperatures, ground states can be used to measure
entanglement through density matrix evaluation, where the non analytical
behavior of the derivative of entanglement explains the phenomenon of QPT 
\cite{xxz, xxzdm, xy, xydm}. Such approaches can be implemented in the QRG
method.

The QRG method was used to solve exactly the one dimensional Ising, XXZ and
XY models \cite{ising, xxz, xxzdm, xy, xydm}. Where it was found that the
nearest neighbors interaction exhibits the QPT near the critical point. For
a deeper insight, the next nearest neighbors interaction was studied in XXZ
model\ \cite{xydm30, xydm31}. The RG method was also used in the one
dimensional Ising and XYZ models in the presence of magnetic field \cite%
{ising, xyz}.The Jordan Wigner transformation was used to solve the Ising
model exactly, where it was found that near critical point, this model
exhibits the maximum value of entanglement for the second nearest neighbors 
\cite{vedral intro(1)}. It was analyzed that in thermodynamic limit the
entanglement of ground state of mutually interacting spin-1/2 particles in a
magnetic field shows cusp like singularities exactly at the critical point 
\cite{vedral intro(1)}. \textit{The QRG method in two dimensional spin
systems is a step forward for the better understanding and answering the
open questions like computational complexity of finding the ground states,
ground state properties, energy spectrum, correlation length, criticality,
quantum phase transition and their connection with entanglement.} Analogous
to Kadanoff's block renormalization group approach in one-dimensional spin
systems \cite{xy30}, we apply to two-dimensional spin systems by dividing
the square lattice of spins into blocks of odd number of spins, which span
the whole lattice.

The rest of the paper is arranged as follow. In Sec. II, we present the
model of the system and describe the mathematical formalism to calculate the
renormalized coupling constant and anisotropic coefficients. The effective
Hamiltonian of the system is obtained in terms of renormalized constants. In
Sec. III we investigate the block-block entanglement and its non analytical
behavior which is related to the QPT. We also study the scaling behavior in
this context. The results are summarized in Sec. IV.

\section{QUANTUM RENORMALIZATION OF XY MODEL IN TWO-DIMENSIONS}

Kadanoff block approach was used in the past to study the QRG method in
one-dimensional spin models \cite{xy3, xy4, xy5, ising, xxz, xxzdm, xy, xydm}%
. In this approach the fixed point is achieved after number of iterations by
virtue of reduction of degrees of freedom.\ We extend this very idea and
implement it on a two-dimensional square lattice of spins, in which the
whole lattice is spanned by square blocks, each consisting of five spins
(FIG. 1), with one spin at the center and four at the corners. Using this
model we obtain the renormalized parameters producing the effective
Hamiltonian similar to the original one. The Hamiltonian of a two
dimensional Heisenberg XY model represented by the square lattice of $%
N\times N$ spins can be written as,

\begin{equation}
H(J,\gamma )=\frac{J}{4}\sum_{i=1}^{N}\sum_{j=1}^{N}((1+\gamma )(\sigma
_{i,j}^{x}\sigma _{i+1,j}^{x}+\sigma _{i,j}^{x}\sigma
_{i,j+1}^{x})+(1-\gamma )(\sigma _{i,j}^{y}\sigma _{i+1,j}^{y}+\sigma
_{i,j}^{y}\sigma _{i,j+1}^{y})),  \label{1}
\end{equation}%
where $J$ is the exchange coupling constant, $\gamma $ is the anisotropy
parameter and $\sigma ^{x},\sigma ^{y}$ are the Pauli matrices. Depending on
the values of $\gamma $ the model reduces to different classes such as $XX$
model for $\gamma =0,$ Ising model for $\gamma =1$ and Ising universality
class for $0<\gamma \leq 1$ \cite{xy22}.

\begin{figure}[ht]
\includegraphics[scale=0.9]{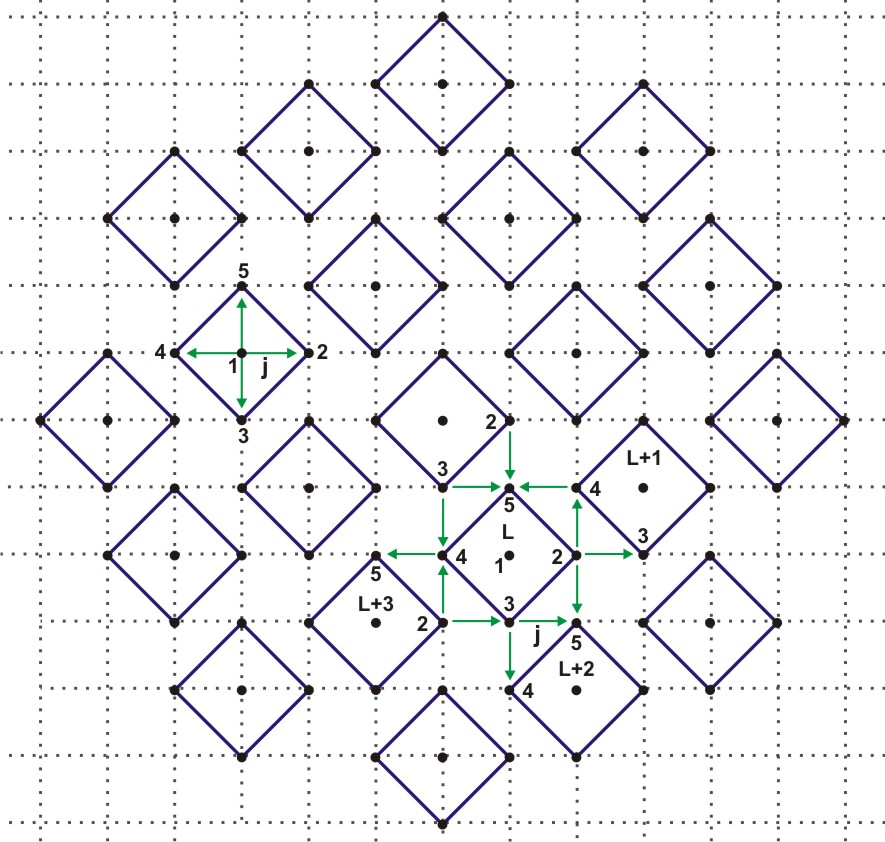}
\caption{(Color online) 2-dimensional square lattice is
depicted by considering each block of five spins.} \label{FIG.1.}
\end{figure} 
We begin by dividing the total Hamiltonian into two parts as

\begin{equation}
H=H^{B}+H^{BB},  \label{2}
\end{equation}%
where $H^{B}$ and $H^{BB}$ are the block and the interblock Hamiltonians
respectively. The explicit form of these Hamiltonians can be written as

\begin{align}
H^{B}& =\frac{J}{4}\sum\limits_{L}^{N/5}((1+\gamma )(\sigma _{L,1}^{x}\sigma
_{L,2}^{x}+\sigma _{L,1}^{x}\sigma _{L,3}^{x}+\sigma _{L,1}^{x}\sigma
_{L,4}^{x}+\sigma _{L,1}^{x}\sigma _{L,5}^{x})  \notag \\
& +(1-\gamma )(\sigma _{L,1}^{y}\sigma _{L,2}^{y}+\sigma _{L,1}^{y}\sigma
_{L,3}^{y}+\sigma _{L,1}^{y}\sigma _{L,4}^{y}+\sigma _{L,1}^{y}\sigma
_{L,5}^{y})),  \label{3}
\end{align}%
and 
\begin{align}
H^{BB}& =\sum\limits_{L}^{N/5}\frac{J}{4}((1+\gamma )(\sigma
_{L,2}^{x}\sigma _{L+1,3}^{x}+\sigma _{L,2}^{x}\sigma _{L+1,4}^{x}+\sigma
_{L,2}^{x}\sigma _{L+2,5}^{x}+\sigma _{L,3}^{x}\sigma _{L+2,4}^{x}  \notag \\
& +\sigma _{L,3}^{x}\sigma _{L+2,5}^{x}+\sigma _{L,4}^{x}\sigma
_{L+3,5}^{x})+(1-\gamma )(\sigma _{L,2}^{y}\sigma _{L+1,3}^{y}+\sigma
_{L,2}^{y}\sigma _{L+1,4}^{y}  \notag \\
& +\sigma _{L,2}^{y}\sigma _{L+2,5}^{y}+\sigma _{L,3}^{y}\sigma
_{L+2,4}^{y}+\sigma _{L,3}^{y}\sigma _{L+2,5}^{y}+\sigma _{L,4}^{y}\sigma
_{L+3,5}^{y})),  \label{4}
\end{align}%
Whereas the $L$th block Hamiltonian can be written as

\begin{align}
H_{L}^{B}& =\frac{J}{4}((1+\gamma )(\sigma _{L,1}^{x}\sigma
_{L,2}^{x}+\sigma _{L,1}^{x}\sigma _{L,3}^{x}+\sigma _{L,1}^{x}\sigma
_{L,4}^{x}+\sigma _{L,1}^{x}\sigma _{L,5}^{x})  \notag \\
& +(1-\gamma )(\sigma _{L,1}^{y}\sigma _{L,2}^{y}+\sigma _{L,1}^{y}\sigma
_{L,3}^{y}+\sigma _{L,1}^{y}\sigma _{L,4}^{y}+\sigma _{L,1}^{y}\sigma
_{L,5}^{y})).  \label{5}
\end{align}%
The interblock interactions are shown by direction of arrows in FIG. 1,
which is mathematically represented by Eq. \ref{4}. We choose block of odd
spins which in turn produces degenerate eigenvalues for the ground state and
makes\ it possible to construct the projection operator in the renamed basis
of the ground state. In terms of matrix product states \cite{xy29}, the
solution i.e., the eigenvalues and the eigenvectors for the single block
Hamiltonian, can be obtained. Therefore, the degenerate lowest energy can be
written as 
\begin{equation}
E_{0}=-\frac{1}{2}J\sqrt{5+5\gamma ^{2}+\alpha _{1}},  \label{6}
\end{equation}%
and the corresponding states in terms of eigenstates $\left\vert \uparrow
\right\rangle $, $\left\vert \downarrow \right\rangle $ of $\sigma ^{z}$ are 
\begin{align}
\left\vert \phi _{0}^{1}\right\rangle & =\gamma _{1}(\left\vert \uparrow
\uparrow \uparrow \uparrow \downarrow \right\rangle +\left\vert \uparrow
\uparrow \uparrow \downarrow \uparrow \right\rangle +\left\vert \uparrow
\uparrow \downarrow \uparrow \uparrow \right\rangle +\left\vert \uparrow
\downarrow \uparrow \uparrow \uparrow \right\rangle )  \notag \\
& +\gamma _{2}(\left\vert \uparrow \uparrow \downarrow \downarrow \downarrow
\right\rangle +\left\vert \uparrow \downarrow \uparrow \downarrow \downarrow
\right\rangle +\left\vert \uparrow \downarrow \downarrow \uparrow \downarrow
\right\rangle +\left\vert \uparrow \downarrow \downarrow \downarrow \uparrow
\right\rangle )  \notag \\
& +\gamma _{3}\left\vert \downarrow \uparrow \uparrow \uparrow \uparrow
\right\rangle +\gamma _{4}(\left\vert \downarrow \uparrow \uparrow
\downarrow \downarrow \right\rangle +\left\vert \downarrow \uparrow
\downarrow \uparrow \downarrow \right\rangle +\left\vert \downarrow \uparrow
\downarrow \downarrow \uparrow \right\rangle  \notag \\
& +\left\vert \downarrow \downarrow \uparrow \uparrow \downarrow
\right\rangle +\left\vert \downarrow \downarrow \uparrow \downarrow \uparrow
\right\rangle +\left\vert \downarrow \downarrow \downarrow \uparrow \uparrow
\right\rangle )+\gamma _{5}\left\vert \downarrow \downarrow \downarrow
\downarrow \downarrow \right\rangle ,  \label{7}
\end{align}%
and 
\begin{align}
\left\vert \phi _{0}^{2}\right\rangle & =\gamma _{6}\left\vert \uparrow
\uparrow \uparrow \uparrow \uparrow \right\rangle +\gamma _{7}(\left\vert
\uparrow \uparrow \uparrow \downarrow \downarrow \right\rangle +\left\vert
\uparrow \uparrow \downarrow \uparrow \downarrow \right\rangle +\left\vert
\uparrow \uparrow \downarrow \downarrow \uparrow \right\rangle  \notag \\
& +\left\vert \uparrow \downarrow \uparrow \uparrow \downarrow \right\rangle
+\left\vert \uparrow \downarrow \uparrow \downarrow \uparrow \right\rangle
+\left\vert \uparrow \downarrow \downarrow \uparrow \uparrow \right\rangle
)+\gamma _{8}\left\vert \uparrow \downarrow \downarrow \downarrow \downarrow
\right\rangle  \notag \\
& +\gamma _{9}(\left\vert \downarrow \uparrow \uparrow \uparrow \downarrow
\right\rangle +\left\vert \downarrow \uparrow \uparrow \downarrow \uparrow
\right\rangle +\left\vert \downarrow \uparrow \downarrow \uparrow \uparrow
\right\rangle +\left\vert \downarrow \downarrow \uparrow \uparrow \uparrow
\right\rangle )  \notag \\
& +\gamma _{10}(\left\vert \downarrow \uparrow \downarrow \downarrow
\downarrow \right\rangle +\left\vert \downarrow \downarrow \uparrow
\downarrow \downarrow \right\rangle +\left\vert \downarrow \downarrow
\downarrow \uparrow \downarrow \right\rangle +\left\vert \downarrow
\downarrow \downarrow \downarrow \uparrow \right\rangle ).  \label{8}
\end{align}%
Expressions for the $\ \alpha _{1},$ and the $\ \gamma _{i}$'$s$ in terms of
the $\gamma $ are given in the appendix.

\begin{figure}[ht]
\includegraphics[scale=0.9]{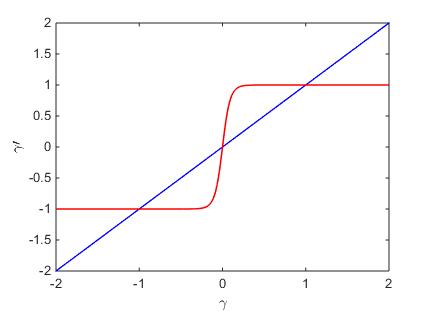}
\caption{(Color online) $\gamma ^{\prime}$ is plotted
against the anisotropic parameter $\gamma $\ for Eq. 17 (red) and
for $\gamma =\gamma ^{\prime}$ (blue). The values of $\protect%
\gamma =0,\pm 1$ provide the solution for $\gamma =\gamma^{\prime}$. }\label{FIG.2.}
\end{figure}

Our aim is to construct the effective Hamiltonian $H^{eff}$ in the
renormalized subspace by finding the renormalized coupling constant and
anisotropy parameter from the projection operators $P_{0}$. For which the
projection operators $P_{0}$ are obtained from the degenerate ground state
eigenvectors of the block Hamiltonian $H^{B}$. The effective Hamiltonian is
related to the original Hamiltonian through \cite{xy30}%
\begin{equation}
H^{eff}=P_{0}^{\dag }HP_{0},  \label{9}
\end{equation}%
where $P_{0}^{\dag }$ is the Hermitian adjoint of $P_{0}$. Using the
perturbative method, we consider only the first order correction term. The
effective Hamiltonian is given by \cite{xxz, xy}

\begin{align}
H^{eff}& =H_{0}^{eff}+H_{1}^{eff}  \notag \\
& =P_{0}^{\dag }H^{B}P_{0}+P_{0}^{\dag }H^{BB}P_{0}.  \label{10}
\end{align}%
In terms of the renamed states of $L$th block, the projection operator $%
P_{0}^{L}$ is defined as \cite{xxz, xy}

\begin{equation}
P_{0}^{L}=\left\vert \Uparrow \right\rangle _{L}\left\langle \phi
_{0}^{1}\right\vert +\left\vert \Downarrow \right\rangle _{L}\left\langle
\phi _{0}^{2}\right\vert ,  \label{11}
\end{equation}%
where $P_{0}$ can be described in product form as%
\begin{equation}
P_{0}=\prod\limits_{L}^{N/5}P_{0}^{L},  \label{12}
\end{equation}%
and $\left\vert \Uparrow \right\rangle _{L}$ and $\left\vert \Downarrow
\right\rangle _{L}$ are the simple qubits of $L$th block to represent
effective site degrees of freedom. The renormalization of the Pauli matrices
is given as

\begin{equation}
P_{0}^{L}\text{ }\sigma _{i,L}^{\varepsilon }\text{ }P_{0}^{L}=\eta
_{i}^{\varepsilon }\text{ }\acute{\sigma}_{L}^{\varepsilon }\text{ \ \ \ \ }%
(i=1,2,3,4,5\text{ };\text{ }\varepsilon =x,y),  \label{13}
\end{equation}%
where%
\begin{eqnarray}
\eta _{1}^{x} &=&4\gamma _{10}\gamma _{2}+\gamma _{3}\gamma _{6}+6\gamma
_{4}\gamma _{7}+\gamma _{5}\gamma _{8}+4\gamma _{1}\gamma _{9},  \notag \\
\eta _{2}^{x} &=&\eta _{3}^{x}=\eta _{4}^{x}=\eta _{5}^{x}  \notag \\
&=&\gamma _{10}(3\gamma _{4}+\gamma _{5})+3\gamma _{2}\gamma _{7}+\gamma
_{1}(\gamma _{6}+3\gamma _{7})+\gamma _{2}\gamma _{8}+\gamma _{9}(\gamma
_{3}-3\gamma _{4}),  \notag \\
\eta _{1}^{y} &=&4\gamma _{10}\gamma _{2}-\gamma _{3}\gamma _{6}-6\gamma
_{4}\gamma _{7}-\gamma _{5}\gamma _{8}+4\gamma _{1}\gamma _{9},  \notag \\
\eta _{2}^{y} &=&\eta _{3}^{y}=\eta _{4}^{y}=\eta _{5}^{y}  \notag \\
&=&\gamma _{10}(3\gamma _{4}-\gamma _{5})-3\gamma _{2}\gamma _{7}+\gamma
_{1}(-\gamma _{6}+3\gamma _{7})+\gamma _{2}\gamma _{8}+\gamma _{9}(\gamma
_{3}+3\gamma _{4}).  \label{14}
\end{eqnarray}

The effective Hamiltonian of the renormalized two dimensional spins surface
is mapped on to the original Hamiltonian with renormalized coupling
parameters, i.e.,

\begin{equation}
H^{eff}=\frac{\acute{J}}{4}\sum_{p=1}^{N/5}\sum_{q=1}^{N/5}((1+\acute{\gamma 
})(\sigma_{p,q}^{x}\sigma_{p+1,q}^{x}+\sigma_{p,q}^{x}\sigma_{p,q+1}^{x})+(1-%
\acute{\gamma})(\sigma_{p,q}^{y}\sigma_{p+1,q}^{y}+\sigma_{p,q}^{y}%
\sigma_{p,q+1}^{y})),  \label{15}
\end{equation}
where%
\begin{align}
\acute{J} & =j(\gamma_{10}^{2}(9\gamma_{4}^{2}+6\gamma\gamma_{4}\gamma
_{5}+\gamma_{5}^{2})+9\gamma_{2}^{2}\gamma_{7}^{2}+\gamma_{1}^{2}(\gamma
_{6}^{2}+6\gamma\gamma_{6}\gamma_{7}+9\gamma_{7}^{2})+6\gamma\gamma_{2}^{2}%
\gamma_{7}\gamma_{8}+\gamma_{2}^{2}\gamma_{8}^{2}  \notag \\
& +6\gamma\gamma_{2}\gamma_{3}\gamma_{7}\gamma_{9}+18\gamma_{2}\gamma
_{4}\gamma_{7}\gamma_{9}+2\gamma_{2}\gamma_{3}\gamma_{8}\gamma_{9}+6\gamma%
\gamma_{2}\gamma_{4}\gamma_{8}\gamma_{9}+\gamma_{3}^{2}\gamma_{9}^{2}+6%
\gamma\gamma_{3}\gamma_{4}\gamma_{9}^{2}  \notag \\
&
+9\gamma_{4}^{2}\gamma_{9}^{2}+2\gamma_{1}(\gamma_{2}(3\gamma_{7}(3\gamma%
\gamma_{7}+\gamma_{8})+\gamma_{6}(3\gamma_{7}+\gamma\gamma
_{8}))+(\gamma\gamma_{3}\gamma_{6}+3\gamma_{4}\gamma_{6}+3\gamma_{3}\gamma
_{7}  \notag \\
& +9\gamma\gamma_{4}\gamma_{7})\gamma_{9})+2\gamma_{10}(\gamma_{1}(\gamma
_{5}\gamma_{6}+9\gamma_{4}\gamma_{7})+\gamma(9\gamma_{2}\gamma_{4}\gamma
_{7}+3\gamma_{1}(\gamma_{4}\gamma_{6}+\gamma_{5}\gamma_{7})  \notag \\
&
+\gamma_{2}\gamma_{5}\gamma_{8}+9\gamma_{4}^{2}\gamma_{9}+\gamma_{3}%
\gamma_{5}\gamma_{9})+3(\gamma_{2}(\gamma_{5}\gamma_{7}+\gamma_{4}\gamma
_{8})+\gamma_{4}(\gamma_{3}+\gamma_{5})\gamma_{9}))),  \label{16}
\end{align}
and

\begin{align}
\acute{\gamma}& =(2(3\gamma _{10}\gamma _{4}+3\gamma _{1}\gamma _{7}+\gamma
_{2}\gamma _{8}+\gamma _{3}\gamma _{9})(\gamma _{10}\gamma _{5}+\gamma
_{1}\gamma _{6}+3\gamma _{2}\gamma _{7}+3\gamma _{4}\gamma _{9})+\gamma
(\gamma _{10}^{2}(9\gamma _{4}^{2}  \notag \\
& +\gamma _{5}^{2})+9\gamma _{2}^{2}\gamma _{7}^{2}+\gamma _{1}^{2}(\gamma
_{6}^{2}+9\gamma _{7}^{2})+\gamma _{2}^{2}\gamma _{8}^{2}+18\gamma
_{2}\gamma _{4}\gamma _{7}\gamma _{9}+2\gamma _{2}\gamma _{3}\gamma
_{8}\gamma _{9}+\gamma _{3}^{2}\gamma _{9}^{2}+9\gamma _{4}^{2}\gamma
_{9}^{2}  \notag \\
& +6\gamma _{1}(\gamma _{2}\gamma _{7}(\gamma _{6}+\gamma _{8})+(\gamma
_{4}\gamma _{6}+\gamma _{3}\gamma _{7})\gamma _{9})+2\gamma _{10}(\gamma
_{1}(\gamma _{5}\gamma _{6}+9\gamma _{4}\gamma _{7})+3(\gamma _{2}(\gamma
_{5}\gamma _{7}  \notag \\
& +\gamma _{4}\gamma _{8})+\gamma _{4}(\gamma _{3}+\gamma _{5})\gamma
_{9}))))/(\gamma _{10}^{2}(9\gamma _{4}^{2}+6\gamma \gamma _{4}\gamma
_{5}+\gamma _{5}^{2})+9\gamma _{2}^{2}\gamma _{7}^{2}+\gamma _{1}^{2}(\gamma
_{6}^{2}+6\gamma \gamma _{6}\gamma _{7}  \notag \\
& +9\gamma _{7}^{2})+6\gamma \gamma _{2}^{2}\gamma _{7}\gamma _{8}+\gamma
_{2}^{2}\gamma _{8}^{2}+6\gamma \gamma _{2}\gamma _{3}\gamma _{7}\gamma
_{9}+18\gamma _{2}\gamma _{4}\gamma _{7}\gamma _{9}+2\gamma _{2}\gamma
_{3}\gamma _{8}\gamma _{9}+6\gamma \gamma _{2}\gamma _{4}\gamma _{8}\gamma
_{9}  \notag \\
& +\gamma _{3}^{2}\gamma _{9}^{2}+6\gamma \gamma _{3}\gamma _{4}\gamma
_{9}^{2}+9\gamma _{4}^{2}\gamma _{9}^{2}+2\gamma _{1}(\gamma _{2}(3\gamma
_{7}(3\gamma \gamma _{7}+\gamma _{8})+\gamma _{6}(3\gamma _{7}+\gamma \gamma
_{8}))+(\gamma \gamma _{3}\gamma _{6}  \notag \\
& +3\gamma _{4}\gamma _{6}+3\gamma _{3}\gamma _{7}+9\gamma \gamma _{4}\gamma
_{7})\gamma _{9})+2\gamma _{10}(\gamma _{1}(\gamma _{5}\gamma _{6}+9\gamma
_{4}\gamma _{7})+\gamma (9\gamma _{2}\gamma _{4}\gamma _{7}+3\gamma
_{1}(\gamma _{4}\gamma _{6}  \notag \\
& +\gamma _{5}\gamma _{7})+\gamma _{2}\gamma _{5}\gamma _{8}+9\gamma
_{4}^{2}\gamma _{9}+\gamma _{3}\gamma _{5}\gamma _{9})+3(\gamma _{2}(\gamma
_{5}\gamma _{7}+\gamma _{4}\gamma _{8})+\gamma _{4}(\gamma _{3}+\gamma
_{5})\gamma _{9}))).  \label{17}
\end{align}

\textit{By solving the Eq}. \ref{17} \textit{for} $\gamma =\acute{\gamma}$, 
\textit{we get the\ solutions} $\gamma =0,\pm 1$ \textit{as shown in FIG. 2}%
. \textit{The model corresponds to the spin fluid phase for} $\gamma
\rightarrow 0$ \textit{which is called the XX model and it corresponds to
Ising like phase for} $\gamma \rightarrow 1$\textit{or} $-1.$ \textit{It
indicates that there lies a phase boundary which separates the two phases.}

\section{STUDY OF ENTANGLEMENT}

We analyze the entanglement by computing the bipartite concurrence of the
interaction between different interblock spins by using the ground state
density matrix. We compute the geometric average of the all possible
bipartite concurrences. The pure density matrix can be written as,

\begin{equation}
\rho =\left\vert \phi _{0}^{1}\right\rangle \left\langle \phi
_{0}^{1}\right\vert ,  \label{18}
\end{equation}%
where $\left\vert \phi _{0}^{1}\right\rangle $ is one of the ground state as
given in Eq. \ref{7}. We calculate the reduced density matrices $\rho
_{23},\rho _{24,}\rho _{25,}\rho _{34,}\rho _{35,}\rho _{45,}$ by taking the
multiple traces and then the bipartite concurrences are worked out. For the
entanglement measurement we compute the geometric mean of all concurrences
through%
\begin{equation}
C_{g}=\sqrt[6]{C_{23}\times C_{24}\times C_{25}\times C_{34}\times
C_{35}\times C_{45}},  \label{19}
\end{equation}%
where $C_{ij}$ $(i,j=2,3,4,5)$ are bipartite concurrences given as \cite%
{xy3132},

\begin{equation}
C_{ij}=\max [\sqrt{\lambda _{ij,4}}-\sqrt{\lambda _{ij,3}}-\sqrt{\lambda
_{ij,2}}-\sqrt{\lambda _{ij,1}},0],  \label{20}
\end{equation}%
where $\lambda _{ij,k}$ for $(k=1,2,3,4)$ are the eigenvalues of \ the
matrix \ $\rho _{ij}\tilde{\rho}_{ij}$ with $\tilde{\rho}_{ij}=(\sigma
_{i}^{y}\otimes \sigma _{j}^{y})$\ $\rho _{ij}^{\ast }(\sigma
_{i}^{y}\otimes \sigma _{j}^{y})$ and $\lambda _{ij,4}>\lambda
_{ij,3}>\lambda _{ij,2}>\lambda _{ij,1}.$

We use the numerical technique to determine the renormalized $\gamma $ and
calculate the average concurrence $C_{g}$ after the each RG iteration. $%
C_{g} $ is plotted against $\gamma $ in FIG. 3 showing its evolution with
increasing the size of the system. The plots of $C_{g}$ coincide with each
other at the critical point. \textit{After two steps (2nd order) }$C_{g}$%
\textit{\ attains two fixed values, (a non-zero value at }$\gamma =0,$%
\textit{\ and zero for }$\gamma \neq 0$\textit{) that predicts the behavior
of the infinitely large system in two dimensions. It indicates that the
two-dimensional surface of spins is effectively equivalent to a five sites
square box with the renormalized coupling constants, thus validating the
idea of the QRG. At }$\gamma =0$\textit{\ the non-zero value of }$C_{g}$%
\textit{\ confirms that system is entangled with no long-range order\ due to
the presence of quantum fluctuations. Such response of the system
corresponds to a spin-fluid phase. For }$\gamma \neq 0$\textit{\ }$(C_{g}=0)$%
\textit{\ the system possesses the magnetic long-range order. Therefore,
nontrivial points i.e., }$\gamma =\pm 1$\textit{\ correspond to two Ising
phases in the }$x$\textit{\ and }$y$\textit{\ directions respectively.} The
results obtained for concurrence in 2D are similar to the one-dimensional
case \cite{xxz, xy}. But the magnitude of concurrence is smaller in 2D,
because the number of shared neighbor sites are larger in 2D as compared to
one-dimensional chain.

\begin{figure}[ht]
\includegraphics[scale=0.9]{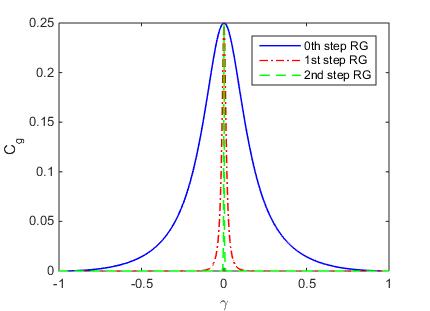}
\caption{(Color online) Geometric average of the
concurrences is plotted against anisotropic parameter $\gamma$
after each step of the RG.}  \label{FIG.3.}
\end{figure}

\textit{The critical behavior of the entanglement can be seen as a diverging
of its derivative when it crosses the phase transition point. The absolute
values of the first derivative of the concurrence with respect to }$\gamma $%
\textit{\ after each iteration are shown in FIG. 4. The diverging behavior
of the derivative at }$\gamma =0$\textit{\ can be seen with increasing the
RG iterations. While concurrence itself remains continuous. It reveals that
the system exhibits the second-order QPT. }It is also noted that the
entanglement in the vicinity of the critical point shows scaling behavior 
\cite{vedral intro(2)}. At the critical point, the entanglement scales
logarithmically and saturates away from the critical point \cite{xxz24}. As
we have discussed earlier a large system $N=5^{n+1},$ can be effectively
represented by five sites box with renormalized coupling constants after the 
$n$th RG iteration. Therefore, the entanglement between the two renormalized
sites describes the entanglement between two blocks, each containing $N/5$
sites. We note that the system shows the scaling behavior which is linear
when $\ln $ of maximum of the absolute value of first derivative $\ln (\mid
dC_{g}/d\gamma \mid _{\max })$\ is plotted against $\ln N=\ln 5^{n+1},$
where $n=1,2,3...$. The scaling behavior is shown in FIG. 5. \textit{The
position of the maximum of }$dC_{g}/d\gamma $\textit{\ approaches the
critical point as the size of the system increases. To get more insight, we
plot }$\ln (\gamma _{c}-\gamma _{\max })$\textit{\ against }$\ln N$\textit{\
in FIG. 6 and obtain the relation }$\gamma _{\max }=\gamma
_{c}-(0.33N)^{-\theta },$\textit{\ where the entanglement exponent }$\theta
=1.14.$\textit{\ The entanglement exponent }$\theta $\textit{\ obtained from
the RG method captures the behavior of the XY model in the vicinity of the
critical point and defined as inverse of the correlation length exponent. In
thermodynamic limit, the correlation length covers the entire system as we
approach the critical point.}

\begin{figure}[ht]
\includegraphics[scale=0.9]{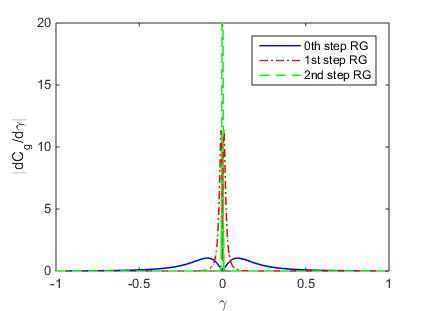}
\caption{(Color online) Absolute derivative of the
geometric average of the concurrences is plotted against $\gamma $
as the RG iteration is increased.}  \label{FIG.4.}
\end{figure}

\begin{figure}[ht]
\includegraphics[scale=0.9]{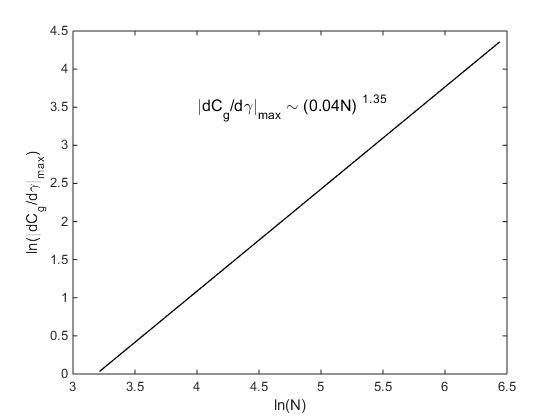}
\caption{Logarithm of the absolute value of
the maximum of the derivative of the concurrence is plotted against the
logarithm of N, the system size.}  \label{FIG.5.}
\end{figure}

\begin{figure}[ht]
\includegraphics[scale=0.9]{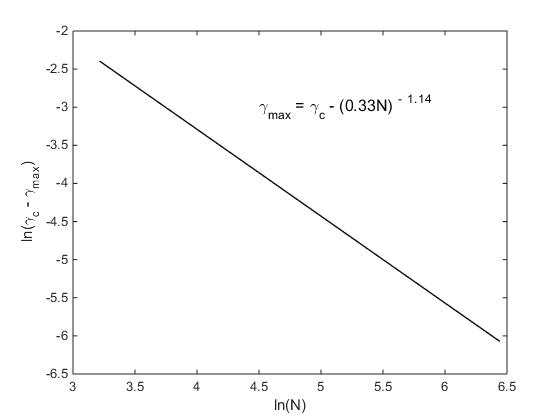}
\caption{Scaling behavior of $\gamma _{\max }$ is plotted against N the size of the system, where $\gamma _{\max }$ is the position of the maximum of the derivative of the
concurrence (see FIG. 6).}  \label{FIG.6.}
\end{figure}

\section{CONCLUSIONS}

Study of the correlated systems in two dimensions through the
renormalization group (RG) technique was presented in this paper. For this
purpose, square lattice of\ Heisenberg spin-1/2 XY model was considered. The
quantum correlations were explored through concurrence and were related to
the quantum phase transition (QPT). Due to the presence of several
interblock interactions, we computed geometric average of the concurrences
of the all possible interactions between the blocks. We noted that the
system size increases rapidly and reaches at the critical point in the less
number of the RG iterations as compared to the one-dimensional case which
were studied previously \cite{xxz, xxzdm, xy, xydm}. \textit{Moreover, we
found that the results for concurrence in 2D are similar to the
one-dimensional case qualitatively. But the magnitude of the concurrence is
smaller in 2D, because the shared neighbor sites are larger in number in 2D
as compared with one-dimensional chain}. The evolution of the entanglement
after the $n$th RG iteration explains that it develops two values, one non
zero value at the critical point and approaches to zero otherwise, which
correspond to spin-fluid phase and Ising phase respectively. The relation
between the critical point, which is maximum value of the absolute
derivative of the concurrence and the system size (scaling behavior) was
investigated, which showed a linear behavior. Moreover, the scaling behavior
was explored through determination of the entanglement exponent which
describes how the critical point is acheived as the size of the system
increases.

\section{ACKNOWLEDGMENTS}

This work was partly supported by the HIGHER EDUCATION COMMISSION, PAKISTAN
under the Indigenous Ph.D. Fellowship Scheme.

\section{APPENDIX}

The expression for $\gamma$'s are given below;

\begin{eqnarray*}
\gamma _{1} &=&-\frac{(-1+\alpha _{1}+\gamma ^{2})\sqrt{(5+\alpha
_{1}+5\gamma ^{2}}}{4\sqrt{2\alpha _{2}}}, \\
\gamma _{2} &=&-\frac{3\sqrt{\frac{\gamma ^{4}(5+\alpha _{1}+5\gamma ^{2})}{%
\alpha _{2}}}}{2\sqrt{2}\gamma }, \\
\gamma _{3} &=&\frac{(-1+\alpha _{1}+\gamma ^{2})}{\sqrt{2\alpha _{2}}}, \\
\gamma _{4} &=&\frac{\gamma (5+\alpha _{1}+\gamma ^{2})}{2\sqrt{2\alpha _{2}%
}}, \\
\gamma _{5} &=&\frac{3\sqrt{2}\gamma ^{2}}{\alpha _{2}}, \\
\gamma _{6} &=&\frac{\sqrt{\frac{\gamma ^{2}(5+\alpha _{1}+5\gamma ^{2})}{%
1+\alpha _{1}+34\gamma ^{2}-\alpha _{1}\gamma ^{2}+\gamma ^{4}}}(-2-2\alpha
_{1}+17\gamma ^{2}-3\alpha _{1}\gamma ^{2}+3\gamma ^{4})}{4(3+2\gamma
^{2}+3\gamma ^{4})}, \\
\gamma _{7} &=&-\frac{\sqrt{\frac{\gamma ^{2}(5+\alpha _{1}+5\gamma ^{2})}{%
1+\alpha _{1}+34\gamma ^{2}-\alpha _{1}\gamma ^{2}+\gamma ^{4}}}(1+\alpha
_{1}-\gamma ^{2}+6\gamma ^{4})}{4\gamma (3+2\gamma ^{2}+3\gamma ^{4})}, \\
\gamma _{8} &=&-\frac{3\sqrt{\frac{\gamma ^{2}(5+\alpha _{1}+5\gamma ^{2})}{%
1+\alpha _{1}+34\gamma ^{2}-\alpha _{1}\gamma ^{2}+\gamma ^{4}}}(5-\alpha
_{1}+5\gamma ^{2})}{4(3+2\gamma ^{2}+3\gamma ^{4})}, \\
\gamma _{9} &=&\frac{(1+\alpha _{1}-\gamma ^{2})}{4\gamma \sqrt{(34-\alpha
_{1}+\frac{1+\alpha _{1}}{\gamma ^{2}}+\gamma ^{2})}}, \\
\gamma _{10} &=&\frac{3}{2\sqrt{(34-\alpha _{1}+\frac{1+\alpha _{1}}{\gamma
^{2}}+\gamma ^{2})}},
\end{eqnarray*}

where,

\begin{eqnarray*}
\alpha _{1} &=&\sqrt{1+34\gamma ^{2}+\gamma ^{4}}, \\
\alpha _{2} &=&2-2\alpha _{1}+71\gamma ^{2}+17\alpha _{1}\gamma
^{2}+104\gamma ^{4}+3\alpha _{1}\gamma ^{4}+3\gamma ^{6}.
\end{eqnarray*}

\end{document}